\documentclass[showpacs,aps,graphicx,twocolumn]{revtex4}
\usepackage{graphicx}

\begin{document}

\title{ Efficient hybrid entanglement concentration for quantum communications}

\author{  Lan Zhou,$^{1,2}$ Yu-Bo Sheng$^{2,3}$\footnote{Email address:
shengyb@njupt.edu.cn} }
\address{
$^1$College of Mathematics \& Physics, Nanjing University of Posts and Telecommunications, Nanjing,
210003, China\\
 $^2$Key Lab of Broadband Wireless Communication and Sensor Network
 Technology,
 Nanjing University of Posts and Telecommunications, Ministry of
 Education, Nanjing, 210003,
 China\\
 $^3$ Institute of Signal Processing  Transmission, Nanjing
University of Posts and Telecommunications, Nanjing, 210003,  China\\}

\begin{abstract}
We present two groups of practical  entanglement concentration
protocols (ECPs) for optical hybrid entangled state (HES).
In the first group, it contains two ECPs and both ECPs do not need to know the initial coefficients of the
less-entangled state. The second group contains three ECPs and they need to know the initial coefficients.
It is shown that the yield of the entanglement concentration in the second group is greater than the first group. Moreover, some protocols are based on
the linear optics and can be easily realized under current experiment condition. These protocols may be
useful in current hybrid quantum communication.
\end{abstract}
 \maketitle

\section{introduction}

The distribution of entanglement over long-distances is essential
for potential future application for quantum communication \cite{rmp,book}, such as
  teleportation \cite{teleportation}, quantum key
distribution(QKD) \cite{Ekert91,BBM92,QKD}, quantum dense
coding \cite{densecoding}, quantum state sharing \cite{QSS1,QSS2,QSS3,QSSpra}, and quantum secure direct communication \cite{long,two-step,lixhpra}. But the
quantum state transformation in quantum channel such as free space
and optical fiber always suffers from the noise. The noise will make the
photon loss and decoherence. It greatly limit the transmission
distance in quantum communication. In order to overcome this flaw, the concept of the
quantum repeater was introduced.
In 1998, Briegel \emph{et al.} proposed the  quantum repeaters
based on quantum purification and quantum
swapping \cite{repeater1,repeater2}. The basic idea is to divide the
whole long-distance channel into many short segments. The
 photons are transmitted  between each segments, which is
comparable to the channel attenuation length.  In 2001, Duan
\emph{et al.} developed this idea and proposed a quantum
repeater  protocol based on the atomic ensembles and optical
elements \cite{DLCZ}. It is usually called the DLCZ protocol. Meanwhile, there are other groups focused on the
quantum repeaters for both theory and
experiments \cite{repeater3,repeater4,repeater5,repeater6,repeater7,repeater8,
repeater9,repeater10,wangtiejun,detail6,jiang,Rydberg5,Rydberg6,repeaterRMP,quantumnetworkrmp}. Most
current schemes  focus on  the heralded creation of very high
fidelity base-level pairs based on the atomic ensembles and linear
optics \cite{repeaterRMP}.

On the other hand, another kind of quantum repeaters named hybrid
quantum repeaters have been widely
discussed \cite{hybid1,hybid2,hybid3,hybid4,hybid5,hybid6,hybid7,hybid8,hybid9,hybid10,hybid11,hybid12,hybrid13,hybridteleportation}.
 Comparing with the previous quantum repeater
protocols, the biggest difference is that in
the stage of entanglement connection, they use the  coherent state
instead of the single photon, to create the hybrid entangled state (HES) as
 \begin{eqnarray}
|\Phi^{+}\rangle=\frac{1}{\sqrt{2}}(|\bar{0}\rangle|\beta\rangle+|\bar{1}\rangle|-\beta\rangle).\label{hybirdbellstate1}
\end{eqnarray}
Here the $|\pm\beta\rangle$ is the coherent states, and
$|\bar{0}\rangle$, $|\bar{1}\rangle$ are the single qubit. The single qubit may be the individual $\Lambda$-type atom,
  the trapped ion,   a neutral donor impurity in semiconductors, the nitrogen-vacancy (NV)
center in a diamond with a nuclear spin,
a single electron trapped in quantum dots \cite{hybid1,hybid3,hybid12}, or a single photon encoded in the polarization
which will make the HES become $\frac{1}{\sqrt{2}}(|H\rangle|\beta\rangle+|V\rangle|-\beta\rangle)$ \cite{hybridteleportation}.  $|H\rangle$ and $|V\rangle$ represent the horizonal and
vertical polarization, respectively.

However, the imperfect operation on the single qubit and coherent state or the
practical noise in the environment may make the maximally HES become a less-entangled state as
\begin{eqnarray}
|\Phi_{\perp}^{+}\rangle=a|\bar{0}\rangle|\beta\rangle+b|\bar{1}\rangle|-\beta\rangle.\label{partialentanglement}
\end{eqnarray}
Here $|a|^{2}+|b|^{2}=1$.
 Unfortunately, the imperfect
entangled state in Eq. (\ref{partialentanglement}) will make the fidelity
of quantum teleportation degrade, the key in quantum cryptography
insecure and quantum dense coding fail. Thus, before entanglement
connection in hybrid quantum repeaters, one should recover the imperfect
entangled states into the maximally ones shown in Eq. (\ref{hybirdbellstate1}).

The entanglement concentration is a powerful way to recover such
less-entangled states into maximally entangled states. It  was  proposed
by Bennett \emph{et al.} \cite{C.H.Bennett2} in 1996, which was
called the Schmidt projection method.  Later some groups showed
that the quantum swapping can also be used to perform the
entanglement concentration \cite{swapping1,swapping2}.  Yamamoto \emph{et al.} \cite{Yamamoto} and Zhao \emph{et
al.} \cite{zhao1} proposed two similar protocols of the concentration with the
linear optics, independently. Both  protocols were
demonstrated by experiments \cite{Yamamoto1,zhao2}. We call them PBS
concentration protocol. The  entanglement concentration
protocols (ECPs)
with cross-Kerr nonlinearity and solid system have also
been proposed \cite{shengpra1,shengpra2,dengsingle,wangchuan1,wangchuan2,wangchuan3,shengreview}.
Unfortunately, current ECPs cannot deal with the HES,
for they   focus on the discrete entangled photon
pairs with the same degree of freedom \cite{C.H.Bennett2,swapping1,swapping2,Yamamoto,zhao1,Yamamoto1,zhao2,shengpra1}.

In this paper, we will present two groups of   ECPs for optical HES. In the first group,
the parties do not need to know the initial coefficients of the less-entangled state. In the second group, they
should know the initial coefficients. Interestingly, it is shown that if they know the
initial coefficients, the total success probability in the second group is much greater than the first group. The  HES in this paper
is encoded in the single photon in polarization and the
coherent state, for the optical HESs   have  been
widely discussed recently \cite{hybridentanglement1,hybridentanglement2,hybridentanglement3,hybridentanglement4,hybridstate1,hybridstate2,hybridstate3,hybridstate5,hybridstate6,hybridstate8}.
Especially, the HES of the form
$\frac{1}{\sqrt{2}}(|H\rangle|\beta\rangle+|V\rangle|-\beta\rangle)$ can be
generated in principle by performing a weak cross-Kerr nonlinear interaction between a single photon and a strong coherent state with
the help of a displacement operation \cite{QND1}, and can be used to perform
a scheme to realize deterministic quantum teleportation \cite{hybridteleportation}. Certainly, the HES encoded in the
other solid qubits and the coherent state can also be concentrated with the similar principle.

The first group contains two ECPs. In the first protocol, we use the polarization beam
splitter (PBS) and beam splitter (BS) to perform a parity check, and
then achieve the concentration task. We call it BS protocol. It is essentially
inspired by the Schmidt projection method \cite{C.H.Bennett2}. The second protocol
 is an improvement of the first one for only one PBS is
needed. We call it BS-improved protocol. The second group contains three ECPs. In the third protocol,
we make a further improvement of the above ECPs.
  In each
concentration step, we use only one pair of hybrid entangled pair
and a single polarized photon, and it can reach the same success probability
as the first one. We call it single-photon protocol.  In the
forth protocol, we resort to the quantum nondemolition (QND) constructed
by cross-Kerr nonlinearity to improve the third protocol. We call it QND
protocol. By repeating this QND protocol, the success probability
can be greatly increased. Moreover, in the five ECP, we do not need any auxiliary photon and it can reach
the same yield as the QND protocol by performing it only one time. Therefore,  it is the optimal one.  All  protocols not only can be
used to concentrate the partially single-photon and single-coherent
state HES shown in Eq. (2), but also can be extended
to deal with the cases of multi-particle and multi-coherent states.

This paper is organized as follows: In Sec. II, we explain the first and the second
protocol, following the same principle of the conventional
ECPs.  In Sec. III, we
describe our third protocol assisted with single photon. The Sec.
IV is the forth protocol constructed by cross-Kerr nonlinearity. In Sec. V, we discuss the optimal ECP
without any auxiliary photons.
Finally, in Sec. VI, we make a discussion and conclusion.

\section{Conventional ECPs with linear optics}
\subsection{Conventional ECPs for single-photon and single-coherent hybrid entangled state}
    Fig. 1 is a schematic drawing
of the basic principle of our ECP. Suppose Alice
and Bob first share two copies of unknown HESs
\begin{eqnarray}
|\Phi_{\perp}^{+}\rangle_{a1b1}=a|H\rangle_{a1}|\beta\rangle_{b1}+b|V\rangle_{a1}|-\beta\rangle_{b1},\label{hybridpartialentangledstate1}
\end{eqnarray}
and
\begin{eqnarray}
|\Phi_{\perp}^{+}\rangle_{a2b2}=a|H\rangle_{a2}|\beta\rangle_{b2}+b|V\rangle_{a2}|-\beta\rangle_{b2}.\label{hybridpartialentangledstate2}
\end{eqnarray}
The single photons are shared by Alice and the coherent states are shared by Bob. The PBS in Alice's location is to transmit the $|H\rangle$ polarized photon
and reflect the $|V\rangle$ polarized photon. The 50:50 BS in Bob's location is to
make
\begin{eqnarray}
|\beta\rangle_{b1}|\beta\rangle_{b2}\longrightarrow|\sqrt{2}\beta\rangle_{d1}|0\rangle_{d2},\label{paritycheck1}
\end{eqnarray}
\begin{eqnarray}
|\beta\rangle_{b1}|-\beta\rangle_{b2}\longrightarrow|0\rangle_{d1}|\sqrt{2}\beta\rangle_{d2},\label{paritycheck2}
\end{eqnarray}
\begin{eqnarray}
|-\beta\rangle_{b1}|\beta\rangle_{b2}\longrightarrow|0\rangle_{d1}|-\sqrt{2}\beta\rangle_{d2},\label{paritycheck3}
\end{eqnarray}
\begin{eqnarray}
|-\beta\rangle_{b1}|-\beta\rangle_{b2}\longrightarrow|-\sqrt{2}\beta\rangle_{d1}|0\rangle_{d2}.\label{paritycheck4}
\end{eqnarray}
Here the $|0\rangle$ is the vacuum state.
Alice first rotates her single photon in $a2$ spatial mode with $90^{\circ}$ by the
half wave plate (HWP). The state $|\Phi_{\perp}^{+}\rangle_{a2b2}$
becomes
\begin{eqnarray}
|\Phi_{\perp}^{+}\rangle'_{a2b2}=a|V\rangle_{a2}|\beta\rangle_{b2}+b|H\rangle_{a2}|-\beta\rangle_{b2}.\label{hybridpartialentangledstate3}
\end{eqnarray}
\begin{center}
\begin{figure}[!h]
\includegraphics[width=7cm,angle=0]{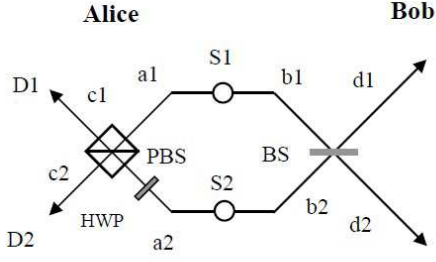}
\caption{Schematic diagram of the proposed HES
concentration. Two pairs of identical less-entangled  HES are sent to Alice and Bob from sources S1 and  S2, respectively. Alice
receives the single photons, and the Bob receives the coherent states.}
\end{figure}
\end{center}
After the states passing though the PBS and BS, the initial system can be rewritten as
\begin{eqnarray}
&&|\Phi_{\perp}^{+}\rangle_{a1b1}\otimes|\Phi_{\perp}^{+}\rangle'_{a2b2}
=(a|H\rangle_{a1}|\beta\rangle_{b1}+b|V\rangle_{a1}|-\beta\rangle_{b1})\nonumber\\
&&\otimes(a|V\rangle_{a2}|\beta\rangle_{b2}+b|H\rangle_{a2}|-\beta\rangle_{b2})\nonumber\\
&&=a^{2}|H\rangle_{a1}|V\rangle_{a2}|\beta\rangle_{b1}|\beta\rangle_{b2}\nonumber\\
&&+b^{2}|V\rangle_{a1}|H\rangle_{a2}|-\beta\rangle_{b1}|-\beta\rangle_{b2}\nonumber\\
&&+ab(|H\rangle_{a1}|H\rangle_{a2}
|\beta\rangle_{b1}|-\beta\rangle_{b2}
+|V\rangle_{a1}|V\rangle_{a2}\nonumber\\
&&|-\beta\rangle_{b1}|\beta\rangle_{b2}) \rightarrow
a^{2}|H\rangle_{c2}|V\rangle_{c2}|\sqrt{2}\beta\rangle_{d1}|0\rangle_{d2}\nonumber\\
&&+b^{2}|V\rangle_{c1}|H\rangle_{c1}|-\sqrt{2}\beta\rangle_{d1}|0\rangle_{d2}
+ab(|H\rangle_{c1}|H\rangle_{c2}|0\rangle_{d1}\nonumber\\
&&|\sqrt{2}\beta\rangle_{d2}
+|V\rangle_{c1}|V\rangle_{c2}|0\rangle_{d1}|-\sqrt{2}\beta\rangle_{d2}).\label{hybridpartialentangledstatecombination}
\end{eqnarray}

From the above equation, the items
$|H\rangle_{c2}|V\rangle_{c2}|\sqrt{2}\beta\rangle_{d1}|0\rangle_{d2}$
and
$|V\rangle_{c1}|H\rangle_{c1}|-\sqrt{2}\beta\rangle_{d1}|0\rangle_{d2}$
will make two photons in the same output mode. But  the other two
items,
$|H\rangle_{c1}|H\rangle_{c2}|0\rangle_{d1}|\sqrt{2}\beta\rangle_{d2}$
$|V\rangle_{c1}|V\rangle_{c2}|0\rangle_{d1}|-\sqrt{2}\beta\rangle_{d2}$
will make the outputs modes $c1$ and $c2$ both contain one photon.
Therefore, by selecting only those events that there is exactly one
photon at each output modes $c1$ and $c2$, Alice and Bob can project
the above state into a maximally HES
\begin{eqnarray}
|\Phi_{\perp}^{+}\rangle_{c}&=&\frac{1}{\sqrt{2}}(|H\rangle_{c1}|H\rangle_{c2}|0\rangle_{d1}|\sqrt{2}\beta\rangle_{d2}\nonumber\\
&+&|V\rangle_{c1}|V\rangle_{c2}|0\rangle_{d1}|-\sqrt{2}\beta\rangle_{d2})),\label{concentrationremained}
\end{eqnarray}
with a success probability of $2|ab|^{2}$.  In order to generate a
maximally entangled state between Alice and Bob, Alice should
measure her photon in $c2$ mode in the basis
$|\pm\rangle=\frac{1}{\sqrt{2}}(|H\rangle\pm|V\rangle)$.  After
performing these operations, if the measurement result is
$|+\rangle$, it will leave the above state as
\begin{eqnarray}
|\Phi_{\perp}^{+}\rangle'_{ab}=\frac{1}{\sqrt{2}}(|H\rangle_{c1}|\sqrt{2}\beta\rangle_{d2}+|V\rangle_{c1}|-\sqrt{2}\beta\rangle_{d2}),\label{generate1}
\end{eqnarray}
 otherwise, it will leave the state as
\begin{eqnarray}
|\Phi_{\perp}^{-}\rangle'_{ab}=\frac{1}{\sqrt{2}}(|H\rangle_{c1}|\sqrt{2}\beta\rangle_{d2}-|V\rangle_{c1}|-\sqrt{2}\beta\rangle_{d2}).\label{generate2}
\end{eqnarray}

\begin{center}
\begin{figure}[!h]
\includegraphics[width=7cm,angle=0]{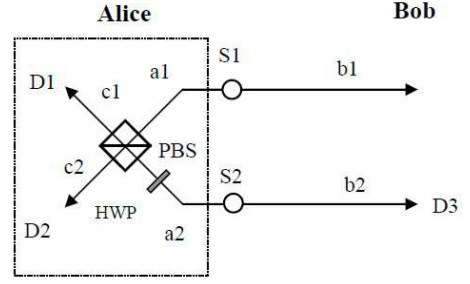}
\caption{Schematic diagram of the improved  HES
concentration protocol. Comparing with Fig. 1, the BS is removed.}
\end{figure}
\end{center}
Both  states are the maximally HESs. In
order to get the same state of $|\Phi_{\perp}^{+}\rangle'_{ab}$,
Alice only needs to perform a simple local operation of phase rotation on her
photon. It is interesting to compare above two states
 with Eq. (\ref{hybirdbellstate1}). The obvious difference is that the amplitude of the
 coherent state is increased. It is quite different from the conventional ECPs.
 In fact, it is an obvious advantage of this protocol, for in a
 practical transmission, the coherent state always suffers from the
 noise, and the photon loss can not be
 avoided \cite{hybid1,hybid2,hybid3,hybid4}. The photon loss will
 decrease the amplitude of the coherent state, while in this protocol,
 after concentration, the amplitude has been increased
 automatically.

 Actually, if they do not need to increase the amplitude of the coherent states, they can improve this ECP as shown in Fig. 2. Comparing with Fig. 1, they remove
 the BS, and make the whole system become
\begin{eqnarray}
&&|\Phi_{\perp}^{+}\rangle_{a1b1}\otimes|\Phi_{\perp}^{+}\rangle'_{a2b2}
=(a|H\rangle_{a1}|\beta\rangle_{b1}+b|V\rangle_{a1}|-\beta\rangle_{b1})\nonumber\\
&&\otimes(a|V\rangle_{a2}|\beta\rangle_{b2}+b|H\rangle_{a2}|-\beta\rangle_{b2})\nonumber\\
&&=a^{2}|H\rangle_{a1}|V\rangle_{a2}|\beta\rangle_{b1}|\beta\rangle_{b2}\nonumber\\
&&+b^{2}|V\rangle_{a1}|H\rangle_{a2}|-\beta\rangle_{b1}|-\beta\rangle_{b2}\nonumber\\
&&+ab(|H\rangle_{a1}|H\rangle_{a2}
|\beta\rangle_{b1}|-\beta\rangle_{b2}
+|V\rangle_{a1}|V\rangle_{a2}\nonumber\\
&&|-\beta\rangle_{b1}|\beta\rangle_{b2}) \rightarrow
a^{2}|H\rangle_{c2}|V\rangle_{c2}|\beta\rangle_{b1}|\beta\rangle_{b2}\nonumber\\
&&+b^{2}|V\rangle_{c1}|H\rangle_{c1}|-\beta\rangle_{b1}|-\beta\rangle_{b2}\nonumber\\
&&+ab(|H\rangle_{c1}|H\rangle_{c2}
|\beta\rangle_{b1}|-\beta\rangle_{b2}
+|V\rangle_{c1}|V\rangle_{c2}\nonumber
|-\beta\rangle_{b1}|\beta\rangle_{b2}).
\end{eqnarray}
Following the same principle, if the spatial modes $c1$ and $c2$ both contain one photon, above state becomes
\begin{eqnarray}
|\Phi_{\perp}^{+}\rangle'_{c}&=&\frac{1}{\sqrt{2}}(|H\rangle_{c1}|H\rangle_{c2}|\beta\rangle_{b1}|-\beta\rangle_{b2}\nonumber\\
&+&|V\rangle_{c1}|V\rangle_{c2}|-\beta\rangle_{b1}|\beta\rangle_{b2})),\label{concentrationremained2}
\end{eqnarray}
with the success probability of $2|ab|^{2}$. In order to obtain the maximally entangled state, Alice measures the
photon in the spatial mode $c2$ in the basis $|\pm\rangle$ and Bob measures the coherent state using  the photon number detector $|n\rangle\langle n|$,
which cannot distinguish the $|\pm\beta\rangle$. To realize the projection $|n\rangle\langle
n|$ deterministically, one should use quantum nondemolition
detection (QND) \cite{he1,lin1}. After performing these measurements, they can obtain
\begin{eqnarray}
|\Phi_{\perp}^{+}\rangle''_{ab}=\frac{1}{\sqrt{2}}(|H\rangle_{c1}|\beta\rangle+|V\rangle|-\beta\rangle_{d2}),\label{generate3}
\end{eqnarray}
if the Alice's measurement result is $|+\rangle$.
 Otherwise, they will obtain
\begin{eqnarray}
|\Phi_{\perp}^{-}\rangle'_{ab}=\frac{1}{\sqrt{2}}(|H\rangle_{c1}|\beta\rangle-|V\rangle|-\beta\rangle_{d2}),\label{generate4}
\end{eqnarray}
if the measurement result is $|-\rangle$.
Both Eqs. (\ref{generate3}) and (\ref{generate4}) are the desired states. In order to obtain Eq. (\ref{generate3}), Alice needs to perform a phase-flip operation on her photon.

The ECPs described in Fig. 1 and Fig. 2 are essentially followed the
traditional ECPs, such as Refs.\cite{zhao1,Yamamoto}. They should require two copies of less-entangled pairs, and do not need
to know the initial coefficients of the states.
 However, the ECPs described in Fig. 1 and Fig. 2 are quite different.  In Fig. 1, both  the single photons and
the coherent states should be operated. After the two single photons being in the different output modes, the two
coherent states passing through BS will make the  amplitude of coherent state increase. Though this ECP cannot obtain the desired maximally
entangled state, the increased amplitude of the coherent state will make this ECP extremely useful in practical quantum communication because the
photon loss. Certainly, if we only require to concentrate the less-entangled state, we can adopt the second ECP by removing the BS.

\subsection{Conventional ECP for multi-photon and multi-coherent state}
Furthermore, it is straightforward   to extend these ECPs to
concentrate the  HES with multi-photon and
multi-coherent state. In this section, we follow the principle of BS protocol to
explain the ECP for multi-photon and multi-coherent state. The BS-improved protocol can also be used
to concentrate the less-entangled state with multi-photon and multi-coherent state. For instance, the initial state is described
as follows
\begin{eqnarray}
|\Phi_{\perp}^{+}\rangle_{NM}&=&a|HH\cdots
H\rangle_{N}|\beta\beta\cdots
\beta\rangle_{M}\nonumber\\
&+&b|VV\cdots V\rangle_{N}|-\beta-\beta\cdots
-\beta\rangle_{M}.\label{NMhybridpartialentangledstate1}
\end{eqnarray}
\begin{figure}[!h]
\includegraphics[width=8cm,angle=0]{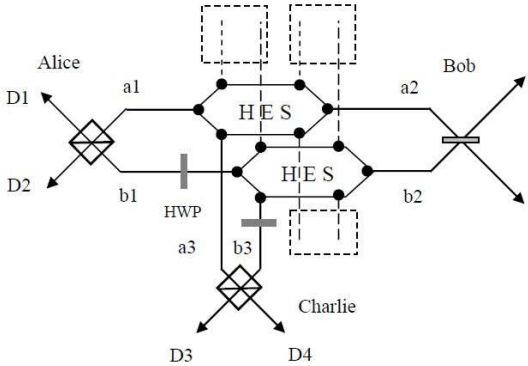}
\caption{Schematic diagram of the ECP for  multi-photon and multi-coherent state HES. Each
parties will receive two single photons or two coherent states.}
\end{figure}
The $N$ is the number of the single photon, and $M$ is the number of
the coherent state. In Fig. 3, two copies of such  states are
distributed to different parties, say Alice, Bob, Charlie, etc.
The whole   composite system can be rewritten as

\begin{eqnarray}
&&|\Phi_{\perp}^{+}\rangle'_{2N2M}=|\Phi_{\perp}^{+}\rangle_{NM}\otimes|\Phi_{\perp}^{+}\rangle_{NM}\nonumber\\
&=&(a|H\rangle_{1}|H\rangle_{2}\cdots |H\rangle_{N}|\beta\rangle_{1}|\beta\rangle_{2}\cdots |\beta\rangle_{M}\nonumber\\
&+&b|V\rangle_{1}|V\rangle_{2}\cdots V\rangle_{N}|-\beta\rangle_{1}|-\beta\rangle_{2}\cdots |-\beta\rangle_{M})\nonumber\\
&\otimes&a|H\rangle_{N+1}|H\rangle_{N+2}\cdots |H\rangle_{2N}|\beta\rangle_{M+1}|\beta\rangle_{M+2}\cdots |\beta\rangle_{2M}\nonumber\\
&+&b|V\rangle_{N+1}|V\rangle_{N+2}\cdots V\rangle_{2N}\nonumber\\
&&|-\beta\rangle_{M+1}|-\beta\rangle_{M+2}\cdots
|-\beta\rangle_{2M}).\label{NMhybridpartialentangledstatecombination1}
\end{eqnarray}
 Alice receives the photon 1 and photon $N+1$, Bob receives
the coherent state $|\beta\rangle_{1}$ and  $|\beta\rangle_{M+1}$,
etc. Before performing the ECP, each parities who own the single photon
first rotate the photons from number $N+1$ to $2N$ by $90^{\circ}$
similar to the case of above section. Therefore, the whole state becomes
\begin{eqnarray}
&&|\Phi_{\perp}^{+}\rangle''_{2N2M}=|\Phi_{\perp}^{+}\rangle_{NM}\otimes|\Phi_{\perp}^{+}\rangle_{NM}\nonumber\\
&=&(a|H\rangle_{1}|H\rangle_{2}\cdots |H\rangle_{N}|\beta\rangle_{1}|\beta\rangle_{2}\cdots |\beta\rangle_{M}\nonumber\\
&+&b|V\rangle_{1}|V\rangle_{2}\cdots V\rangle_{N}|-\beta\rangle_{1}|-\beta\rangle_{2}\cdots |-\beta\rangle_{M})\nonumber\\
&\otimes&a|V\rangle_{N+1}|V\rangle_{N+2}\cdots |V\rangle_{2N}|\beta\rangle_{M+1}|\beta\rangle_{M+2}\cdots |\beta\rangle_{2M}\nonumber\\
&+&b|H\rangle_{N+1}|H\rangle_{N+2}\cdots H\rangle_{2N}\nonumber\\
&&|-\beta\rangle_{M+1}|-\beta\rangle_{M+2}\cdots
|-\beta\rangle_{2M}).\label{NMhybridpartialentangledstatecombination2}
\end{eqnarray}
From Fig. 3,  each single photon  passes through the PBSs
and each coherent state passes through the BSs. Finally, they choose
the cases that each detector after PBS in Alice's location exactly
registers one photon, and they will get
\begin{eqnarray}
|\Phi_{\perp}^{+}\rangle'''_{2N2M}&=&\frac{1}{2}(|H\rangle_{1}|H\rangle_{2}\cdots
|H\rangle_{2N})|0\rangle_{1}|0\rangle_{2}\cdots |0\rangle_{M}\nonumber\\
&&|\sqrt{2}\beta\rangle_{M+1}|\sqrt{2}\beta\rangle_{M+2}\cdots
|\sqrt{2}\beta\rangle_{2M}\nonumber\\
&+&|V\rangle_{1}|V\rangle_{2}\cdots
|V\rangle_{2N})|0\rangle_{1}|0\rangle_{2}\cdots
|0\rangle_{M}\nonumber\\
&&|-\sqrt{2}\beta\rangle_{M+1}|-\sqrt{2}\beta\rangle_{M+2}\cdots
|-\sqrt{2}\beta\rangle_{2M},\nonumber\\
\end{eqnarray}
with the probability of $2|ab|^{2}$. The above state is
also the maximally entangled state. By measuring the photons from
$N+1$ to $2N$ in $|\pm\rangle$ basis, the state of the composite
system becomes
\begin{eqnarray}
&&|\Phi_{\perp}^{+}\rangle'''_{2N2M}\nonumber\\
&&=\frac{1}{2}[|H\rangle_{1}|H\rangle_{2}\cdots
|H\rangle_{N}(\frac{1}{\sqrt{2}})^{\otimes
N}(|H\rangle+|V\rangle)^{\otimes N} \nonumber\\
&&|0\rangle_{1}|0\rangle_{2}\cdots |0\rangle_{M}
|\sqrt{2}\beta\rangle_{M+1}|\sqrt{2}\beta\rangle_{M+2}\cdots
|\sqrt{2}\beta\rangle_{2M}\nonumber\\
&&+|V\rangle_{1}|V\rangle_{2}\cdots
|V\rangle_{N}(\frac{1}{\sqrt{2}})^{\otimes
N}(|H\rangle-|V\rangle)^{\otimes N}\nonumber\\
&&|0\rangle_{1}|0\rangle_{2}\cdots |0\rangle_{M}\nonumber\\
&&|-\sqrt{2}\beta\rangle_{M+1}|-\sqrt{2}\beta\rangle_{M+2}\cdots
|-\sqrt{2}\beta\rangle_{2M}].
\end{eqnarray}
If the number of the single-photon  outcome in $|V\rangle$
is even, they will get
\begin{eqnarray}
&&|\Phi_{\perp}^{+}\rangle'''_{NM}=\frac{1}{\sqrt{2}}(|H\rangle_{1}|H\rangle_{2}\cdots
|H\rangle_{N} \nonumber\\
&& |\sqrt{2}\beta\rangle_{1}|\sqrt{2}\beta\rangle_{2}\cdots
|\sqrt{2}\beta\rangle_{M}\nonumber\\
&&+|V\rangle_{1}|V\rangle_{2}\cdots |V\rangle_{N}\nonumber\\
 &&|-\sqrt{2}\beta\rangle_{M}|-\sqrt{2}\beta\rangle_{M}\cdots
|-\sqrt{2}\beta\rangle_{M}),\label{mulmax1}
\end{eqnarray}
otherwise, they will get
\begin{eqnarray}
&&|\Phi_{\perp}^{-}\rangle'''_{NM}=\frac{1}{\sqrt{2}}(|H\rangle_{1}|H\rangle_{2}\cdots
|H\rangle_{N} \nonumber\\
&& |\sqrt{2}\beta\rangle_{1}|\sqrt{2}\beta\rangle_{2}\cdots
|\sqrt{2}\beta\rangle_{M}\nonumber\\
&&-|V\rangle_{1}|V\rangle_{2}\cdots |V\rangle_{N}\nonumber\\
&& |-\sqrt{2}\beta\rangle_{M}|-\sqrt{2}\beta\rangle_{M}\cdots
|-\sqrt{2}\beta\rangle_{M}).\label{mulmax2}
\end{eqnarray}
Both states in Eqs. (\ref{mulmax1}) and (\ref{mulmax2}) are the maximally entangled states.
In order to obtain $|\Phi_{\perp}^{+}\rangle'''_{NM}$, they should perform a phase-flip operation on one
of the single polarized photons to convert $|\Phi_{\perp}^{-}\rangle'''_{NM}$ to $|\Phi_{\perp}^{+}\rangle'''_{NM}$.

\section{Hybrid ECP assisted with single photon}

In above section, we have explained two ECPs for HES. One of the advantages of such ECPs
is that they do not need to know the exact coefficients of the
initial less-entangled state, for they always select the same copies.
It is not difficult for Alice and Bob to know
the information about the parameters $a$ and $b$ if they can measure
an enough number of sample photon pairs, during a practical quantum
communication \cite{dengsingle}.
 Actually, if the initial coefficients $a$ and $b$ is exactly known, above ECPs will be further simplified and improved.
 That is,
they only need one single photon to complete the task. Inspirited by the Ref. \cite{shengpra2},
the basic principle of this improved ECP is shown in Fig. 4.   The S1 emits the state similar to
Eq. (\ref{hybridpartialentangledstate1}), and S2 emits a
single photon of the form
\begin{eqnarray}
|\Phi\rangle_{a2}=a|H\rangle_{a2}+b|V\rangle_{a2}.\label{singlephoton}
\end{eqnarray}
Alice first rotates the $|\Phi\rangle_{a2}$ by HWP and makes it
become
\begin{eqnarray}
|\Phi\rangle_{a3}=a|V\rangle_{a3}+b|H\rangle_{a3}.
\end{eqnarray}
\begin{figure}[!h]
\includegraphics[width=7cm,angle=0]{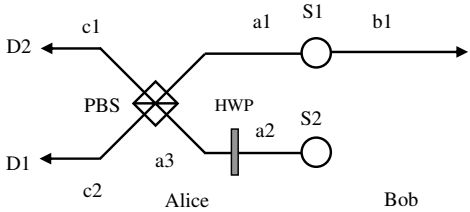}
\caption{Schematic diagram of the concentration protocol assisted with single photons. Comparing with Fig. 1, the most difference is
that the source of S2 emits a single photon. This figure is rather similar to Ref.\cite{shengpra2}.
The difference is that the S1 emits the less-entangled HES.}
\end{figure}
Then the whole system $|\Phi_{\perp}^{+}\rangle_{a1b1}$ combined
with $|\Phi\rangle_{a3}$ evolves as
\begin{eqnarray}
&&|\Phi_{\perp}^{+}\rangle_{a1b1}\otimes|\Phi\rangle_{a3}\nonumber\\
&&=(a|H\rangle_{a1}|\beta\rangle_{b1}+b|V\rangle_{a1}|-\beta\rangle_{b1})
\otimes(a|V\rangle_{a3}+b|H\rangle_{a3})\nonumber\\
&&=a^{2}|H\rangle_{a1}|V\rangle_{a3}|\beta\rangle_{b1}
+b^{2}|V\rangle_{a1}|H\rangle_{a3}|-\beta\rangle_{b1}\nonumber\\
&&+ab(|H\rangle_{a1}|H\rangle_{a3}|\beta\rangle_{b1}+|V\rangle_{a1}|V\rangle_{a3}|-\beta\rangle_{b1}).
\end{eqnarray}
After the PBS, it is easy to find that the items
$|H\rangle|V\rangle|\beta\rangle$ and
$|V\rangle|H\rangle|-\beta\rangle$ will make the two photons in the
same output mode. Only $|H\rangle|H\rangle|\beta\rangle$ and
$|V\rangle|V\rangle|-\beta\rangle$ make the both output modes
contain one photon. By selecting the case that both spatial mode $c1$ and $c2$ containing one photon, the above state collapses to
\begin{eqnarray}
|\Phi_{\bot}^{+}\rangle_{1}=\frac{1}{\sqrt{2}}(|H\rangle_{c1}|H\rangle_{c2}|\beta\rangle_{b1}+|V\rangle_{c1}|V\rangle_{c2}|-\beta\rangle_{b1}),
\end{eqnarray}
with the probability of $2|ab|^{2}$. Then Alice measures the photon
in $c2$ mode in the basis
$|\pm\rangle=\frac{1}{\sqrt{2}}(|H\rangle\pm|V\rangle)$. If the
measurement result is $|+\rangle$, they will get
\begin{eqnarray}
|\Phi_{\bot}^{+}\rangle''_{ab}=\frac{1}{\sqrt{2}}(|H\rangle_{c1}|\beta\rangle_{b1}+|V\rangle_{c1}|-\beta\rangle_{b1}),
\end{eqnarray}
 otherwise, if the measurement result is $|-\rangle$, they will get
\begin{eqnarray}
|\Phi_{\bot}^{-}\rangle''_{ab}=\frac{1}{\sqrt{2}}(|H\rangle_{c1}|\beta\rangle_{b1}-|V\rangle_{c1}|-\beta\rangle_{b1}).
\end{eqnarray}
Both states are the maximally HESs. In
order to get the same state of $|\Phi_{\perp}^{+}\rangle''_{ab}$,
Alice only performs a simple local operation of phase rotation on her
photon.
\begin{figure}[!h]
\includegraphics[width=8cm,angle=0]{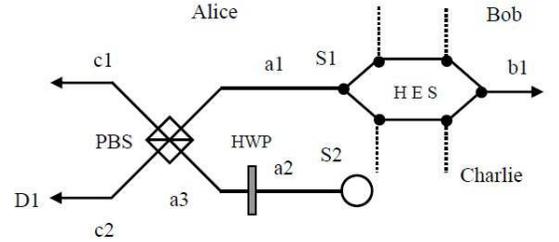}
\caption{Schematic diagram of the concentration protocol for
multi-photon and multi-coherent state, assisted with single photon.
It is much simpler than Fig. 3, for each parties only receive a single
photon or a coherent state.}
\end{figure}

It is straightforward to extend this protocol to the case of
multi-photon and multi-coherent state. The basic principle is shown
in Fig. 5. The source S1 still distributes the HES
with the form of Eq. (\ref{NMhybridpartialentangledstate1}) to each
parties. The source S2 prepares a single photon with the same form
of Eq. (\ref{singlephoton}). After performing a bit-flip on the
Eq. (\ref{singlephoton}), the whole system can be written as
\begin{eqnarray}
&&|\Phi_{\perp}^{+}\rangle_{NM}\otimes|\Phi\rangle_{a3}=(a|H\rangle|H\cdots
H\rangle_{N}|\beta\beta\cdots
\beta\rangle_{M}\nonumber\\
&&+b|V\rangle|V\cdots V\rangle_{N}|-\beta-\beta\cdots
-\beta\rangle_{M}) \otimes (a|V\rangle+b|H\rangle)\nonumber\\
&&=a^{2}|H\rangle|HV\rangle|H\cdots H\rangle_{N+1}|\beta\beta\cdots
\beta\rangle_{M}\nonumber\\
&&+b^{2}|V\rangle|H\rangle|V\cdots V\rangle_{N+1}|-\beta-\beta\cdots
-\beta\rangle_{M}\nonumber\\
&&+ab(|H\rangle|H\cdots H\rangle_{N+1}|\beta\beta\cdots
\beta\rangle_{M}\nonumber\\
&&+|V\rangle|V\cdots V\rangle_{N+1}|-\beta-\beta\cdots
-\beta\rangle_{M}).
\end{eqnarray}
In above equation, for simple, we omit the spatial modes $a1$,
$a3$, $b1$ etc. After passing through the PBS, if the spatial mode $c1$ and $c2$ both contain one photon,
  the combined state in
above equation will collapse to
\begin{eqnarray}
&&|\Phi_{\perp}^{+}\rangle'_{N+1M}=\frac{1}{\sqrt{2}}(|H\rangle|H\cdots
H\rangle_{N+1}|\beta\beta\cdots
\beta\rangle_{M}\nonumber\\
&&+|V\rangle|V\cdots V\rangle_{N+1}|-\beta-\beta\cdots
-\beta\rangle_{M}),
\end{eqnarray}
with the probability of $2|ab|^{2}$. Following the same principle,
Alice measures his photon in $a3$ mode in the basis
$|\pm\rangle$. If the measurement result is $|+\rangle$,
they will get
\begin{eqnarray}
&&|\Phi_{\perp}^{+}\rangle'_{NM}=\frac{1}{\sqrt{2}}(|H\rangle|H\cdots
H\rangle_{N}|\beta\beta\cdots
\beta\rangle_{M}\nonumber\\
&&+|V\rangle|V\cdots V\rangle_{N}|-\beta-\beta\cdots
-\beta\rangle_{M}),
\end{eqnarray}
otherwise, they will get
\begin{eqnarray}
&&|\Phi_{\perp}^{-}\rangle'_{NM}=\frac{1}{\sqrt{2}}(|H\rangle|H\cdots
H\rangle_{N}|\beta\beta\cdots
\beta\rangle_{M}\nonumber\\
&&+|V\rangle|V\cdots V\rangle_{N}|-\beta-\beta\cdots
-\beta\rangle_{M}).
\end{eqnarray}
From above discussion, the ECP assisted with the single photon make the whole
protocol rather simple. On one hand, in each concentration step, they only require one pair of less-entangled state
and can reach the same success probability with the first one, so that ECP is more economical than the first one.
On the other hand, in the conventional ECPs, both Alice and Bob need
to operate or measure their photons. But in the improved ECP, Bob needs to do nothing
but to remain or discard his photons according to the Alice's measurement results.
This feature is rather useful for concentrating the multi-photon and multi-coherent state. In the BS protocol, each parties who owns the  coherent states should
let his two coherent states pass though the BS, while in the BS-improved protocol, they should measure one of their coherent states using the
photon number projection. Therefore, in the single-photon protocol, they can reduce the operation and make the whole concentration step simple.

\section{Hybrid ECP with cross-Kerr nonlinearity}
In Sec. III, we have described our ECP with linear optics. It is easy
to realize it in current technology. However, it is not optimal. The
total success probability is only $2|ab|^{2}$. On the other hand, it
is based on the post-selection principle. That is to say, after
concentration, the photon will be destroyed. In this section, we
will adopt QND constructed by weak cross-Kerr
nonlinearity to achieve this task \cite{QND1,QND2}. After performing the
concentration, the success probability can be greatly increased by
repeating this protocol.
\begin{figure}[!h]
\includegraphics[width=8cm,angle=0]{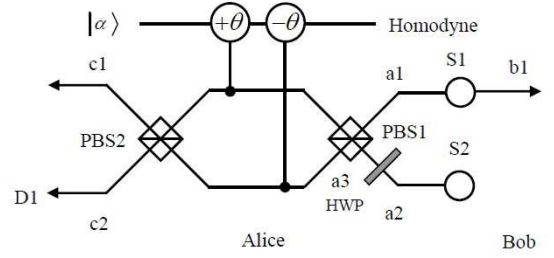}
\caption{Schematic diagram of the ECP for HES with cross-Kerr nonlinearity.}
\end{figure}
From Fig. 6, the  Hamiltonian of a cross-Kerr nonlinear medium can be
written as $H=\hbar\chi \hat{n_{a}}\hat{n_{b}}$, where the
$\hbar\chi$ is the coupling strength of the
nonlinearity \cite{QND1,QND2}. It is decided by the material of
cross-Kerr. Cross-Kerr nonlinearity has been widely studied in
quantum information
processing \cite{QND1,QND2,shengpra,he1,lin1,shengpra1,qi,
dengsingle}.
The basis principle can be described as follows: if the coherent
state $|\alpha\rangle$ combined with a quantum state
$|\varphi\rangle$ couples with the cross-Kerr nonlinearity, the
coherent state $|\alpha\rangle$ will pick up a phase shift. The
phase shift is proportional to the photon number of the quantum
state $|\varphi\rangle$. If the photon number of
$|\varphi\rangle$ is $n$, the coherent state evolves to $|\alpha
e^{in\theta}\rangle$. $\theta=\chi t$ and  $t$ is the interaction
time.

Now we reconsider the system
$|\Phi_{\perp}^{+}\rangle_{a1b1}\otimes|\Phi\rangle_{a3}$ combined
with the coherent state $|\alpha\rangle$
\begin{eqnarray}
&&|\Phi_{\perp}^{+}\rangle_{a1b1}\otimes|\Phi\rangle_{a3}\otimes|\alpha\rangle\nonumber\\
&&=(a|H\rangle_{a1}|\beta\rangle_{b1}+b|V\rangle_{a1}|-\beta\rangle_{b1})
\otimes(a|V\rangle_{a3}+b|H\rangle_{a3})\nonumber\\
&&\otimes|\alpha\rangle\rightarrow
a^{2}|H\rangle_{a1}|V\rangle_{a3}|\beta\rangle_{b1}|\alpha
e^{-i2\theta}\rangle\nonumber\\
&&+b^{2}|V\rangle_{a1}|H\rangle_{a3}|-\beta\rangle_{b1}|\alpha
e^{i2\theta}\rangle\nonumber\\
&&+ab(|H\rangle_{a1}|H\rangle_{a3}|\beta\rangle_{b1}+|V\rangle_{a1}|V\rangle_{a3}|-\beta\rangle_{b1})|\alpha\rangle.
\end{eqnarray}
Obviously, if $|\alpha\rangle$ picks up no phase shift, the system
will collapse to the state $|\Phi_{\bot}^{+}\rangle_{1}$, with the
probability of $2|ab|^{2}$. It is the same as the case of using
linear optics. Following the same principle described in Sec. III,
after measuring the photon in $c2$ spatial mode in the basis $|\pm\rangle$, they will get
$|\Phi_{\bot}^{+}\rangle''_{ab}$, if the measurement result is $|+\rangle$, or get $|\Phi_{\bot}^{-}\rangle''_{ab}$ if
 the measurement result is $|-\rangle$. However, there is another case that, if the
$|\alpha\rangle$ picks up the phase shift $2\theta$. Here the
homodyne measurement can make the $\pm2\theta$
undistinguished. Then the remained state is
\begin{eqnarray}
|\Phi_{\bot}^{+}\rangle'_{1}=a^{2}|H\rangle_{a1}|V\rangle_{a3}|\beta\rangle_{b1}+b^{2}|V\rangle_{a1}|H\rangle_{a3}|-\beta\rangle_{b1}.
\end{eqnarray}
Then after measuring the photon in $a3$ spatial mode in $|\pm\rangle$, it becomes
\begin{eqnarray}
|\Phi_{\bot}^{\pm}\rangle''_{1}&=&\frac{a^{2}}{\sqrt{|a|^{4}+|b|^{2}}}|H\rangle_{a1}|\beta\rangle_{b1}\nonumber\\
&\pm&
\frac{b^{2}}{\sqrt{|a|^{4}+|b|^{4}}}|V\rangle_{a1}|-\beta\rangle_{b1}.\label{lessentangled}
\end{eqnarray}
Both  $|\Phi_{\bot}^{\pm}\rangle''_{1}$ are less-entangled states,
which can be reconcentrated in the next step. We take
$|\Phi_{\bot}^{+}\rangle''_{1}$ for example. Source S2 emits
another single photon with the form of
\begin{eqnarray}
|\Phi\rangle'_{a2}=\frac{a^{2}}{\sqrt{|a|^{4}+|b|^{4}}}|H\rangle_{a2}+\frac{b^{4}}{\sqrt{|a|^{4}+|b|^{4}}}|V\rangle_{a2}.
\end{eqnarray}
After a bit-flip operation, it becomes
\begin{eqnarray}
|\Phi\rangle'_{a3}=\frac{a^{2}}{\sqrt{|a|^{4}+|b|^{4}}}|V\rangle_{a3}+\frac{b^{4}}{\sqrt{|a|^{4}+|b|^{4}}}|H\rangle_{a3}.
\end{eqnarray}
Then the $|\Phi_{\bot}^{+}\rangle''_{1}\otimes|\Phi\rangle'_{a3}$
combined with the $|\alpha\rangle$ evolves as
\begin{eqnarray}
&&|\Phi_{\bot}^{+}\rangle''_{1}\otimes|\Phi\rangle'_{a3}\otimes|\alpha\rangle=(\frac{a^{2}}{\sqrt{|a|^{4}+|b|^{4}}}|H\rangle_{a1}|\beta\rangle_{b1}\nonumber\\
&&+
\frac{b^{2}}{\sqrt{|a|^{4}+|b|^{4}}}|V\rangle_{a1}|-\beta\rangle_{b1})\nonumber\\
&&\otimes(\frac{a^{2}}{\sqrt{|a|^{4}+|b|^{4}}}|V\rangle_{a3}+\frac{b^{2}}{\sqrt{|a|^{4}+|b|^{4}}}|H\rangle_{a3})\otimes|\alpha\rangle\nonumber\\
&&\rightarrow\frac{a^{4}}{|a|^{4}+|b|^{4}}|H\rangle_{a1}|V\rangle_{a3}|\beta\rangle_{b1}|\alpha
e^{-i2\theta}\rangle\nonumber\\
&&+\frac{b^{4}}{|a|^{4}+|b|^{4}}|V\rangle_{a1}|H\rangle_{a3}|-\beta\rangle_{b1}|\alpha
e^{i2\theta}\rangle\nonumber\\
&&+\frac{a^{2}b^{2}}{|a|^{4}+|b|^{4}}(|H\rangle_{a1}|H\rangle_{a3}|\beta\rangle_{b1}+|V\rangle_{a1}|V\rangle_{a3}|-\beta\rangle_{b1})|\alpha\rangle.\nonumber\\
\end{eqnarray}
Similarly, if $|\alpha\rangle$ picks up no phase shift, the above
state can also collapse to the maximally entangled state with the
form of $|\Phi_{\perp}^{+}\rangle_{1}$, with the success probability
of  $\frac{2|ab|^{4}}{|a|^{4}+|b|^{4}}$. The remained state can also
be used to perform further concentration in the next step. It
has the same success probability as the protocol in
Ref. \cite{shengpra1}. In addition, it can also be extended to the case
of multi-partite and multi-coherent HES. Briefly
speaking, after the multi-photon and multi-coherent state
distributed to each parities, one needs to substitute the PBS in
Fig. 4 to the QND in Fig. 6, in Alice's location.

\section{Hybrid ECP without any auxiliary photons}
Recently, the group of Deng present a practical
hyperentanglement concentration for two-photon
four-qubit systems with linear optics \cite{hyperconcentration}.
In their protocol, it is shown that the arbitrary
two-photon less-entangled state can be concentrated
without any assisted photons, if they know
exact coefficients of the initial state.
Interestingly, their excellent idea is suitable for this hybrid
ECP. Here we call it VBS protocol. The VBS is the variable beam splitter with the reflection coefficient $R=a/b$. As shown in Fig. 7, suppose that the Alice receive the single photon and
Bob receive the coherent state. Suppose they know the coefficients $a$ and
$b$, with $b>a$ and $|a|^{2}+|b|^{2}=1$. Therefore, after the state passing through the PBS1, VBS, and PBS2, it evolves as
 \begin{eqnarray}
|\Phi_{\perp}^{+}\rangle_{a1b1}&=&a|H\rangle_{a1}|\beta\rangle_{b1}+b|V\rangle_{a1}|-\beta\rangle_{b1}\nonumber\\
&\rightarrow& a|H\rangle_{a1}|\beta\rangle_{b1}+b|V\rangle_{c1}|-\beta\rangle_{b1}\nonumber\\
&\rightarrow& a|H\rangle_{a1}|\beta\rangle_{b1}+a|V\rangle_{c2}|-\beta\rangle_{b1}\nonumber\\
&+&\sqrt{|b|^{2}-|a|^{2}}|V\rangle_{c3}|-\beta\rangle_{b1}\nonumber\\
&\rightarrow& a(|H\rangle_{a3}|\beta\rangle_{b1}+a|V\rangle_{a3}|-\beta\rangle_{b1})\nonumber\\
&+&\sqrt{|b|^{2}-|a|^{2}}|V\rangle_{c3}|-\beta\rangle_{b1}.\label{without}
\end{eqnarray}
\begin{figure}[!h]
\includegraphics[width=8cm,angle=0]{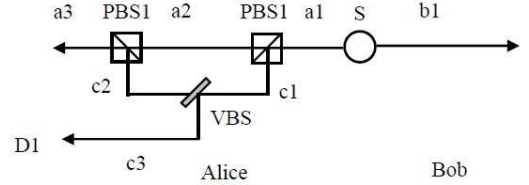}
\caption{Schematic diagram of the ECP without any auxiliary photons.}
\end{figure}
From Eq. (\ref{without}), if the single photon detector D1 detects the single photon, above state will collapse
to the term $|V\rangle_{c3}|-\beta\rangle_{b1}$. On the other hand, if the single photon detector does not click any
photon,  the above state will collapse to the maximally entangled state with the success probability $2|a|^{2}$.
Certainly, if $|a|>|b|$, this ECP can also work by adding the HWP in the spatial mode $a1$.

\section{discussion and summary}
By far, we have fully described our hybrid ECPs. It is interesting
to compare these protocols with the conventional ECPs \cite{zhao1,Yamamoto1}. In BS
protocol, it is essential to follow the similar idea as the
conventional ones. In each step, one has to choose two copies of
less-entangled pairs. After performing the protocol, at least one pair
can be remained. The PBS is used to perform the parity check for the
single photons and the BS acts the same role as the PBS, but for
coherent states.  After picking up the even parity states, the BS
will make one output mode contain no photons. In another output
mode, the amplitude of the coherent state has been increased. The
increased coherent state become a great advantage because of the photon
loss during the transmission. In single-photon protocol, we use the
single photon to concentrate the HES, and reach
the same success probability with the BS protocol. This protocol is
rather simple. Only one PBS and two conventional detectors are
required. Moreover, in each step, we only need one pair of
less-entangled pair but can reach the same success probability as
the BS protocol, which is more commercial than  BS protocol.   In
addition, only one parties say Alice needs to operate this protocol.
This feature makes it more powerful if consider the case of
multi-photon and multi-coherent state. In BS protocol, each of
parties should perform the parity check using PBS or BS. After that,
they should exchange their measurement results to judge the whole
process whether it is a success or failure by classical communication.
Interestingly, in single-photon protocol, after Alice performing the parity
check, she will ask other parties to remain or discard their photons.
That is to say, here only one-way classical communication is
required. The QND protocol is an improvement of the
single-photon protocol. By introducing the QND to substitute the
PBS, the whole protocol can be repeated to get a higher success
probability.   After successfully performing the ECP, the maximally
HES can be remained to perform further
application. In the last ECP, the whole ECP does not require
any auxiliary photon, and can reach the same success probability
as the QND protocol with only linear optics.  We can calculate the yield of the maximally entangled state obtained
in each protocols. We denote the yield as
\begin{eqnarray}
Y_{1}=\frac{N_{c}}{N_{b}}.
\end{eqnarray}
Here $N_{b}$ is the number of originally less-entangled pairs and
$N_{c}$ is the number of maximally entangled pairs after concentration.
Obviously, in BS protocol and BS-improved protocol, they are
\begin{eqnarray}
Y_{1}=|ab|^{2},
\end{eqnarray}
while in single-photon protocol, it is equal to
\begin{eqnarray}
Y'_{1}=2|ab|^{2}.
\end{eqnarray}
In QND protocol, we can get
\begin{eqnarray}
Y''_{1}&=&2|ab|^{2},\nonumber\\
Y''_{2}&=&(1-2|ab|^{2})(\frac{2|ab|^{4}}{(|a|^{4}+|b|^{4})^{2}})=\frac{2|ab|^{4}}{|a|^{4}+|b|^{4}}\nonumber\\
\cdots \nonumber\\
Y''_{K}&=&\frac{2|ab|^{2^{K}}}{(|a|^{4}+|b|^{4})(|a|^{8}+|b|^{8})\cdots(|a|^{2^{N}}+|b|^{2^{K}})}\nonumber\\
&=&\prod^{\infty}_{K=1} \frac{2|ab|^{2^{K}}}{(|a|^{2^{K}}+|b|^{2^{K}})}
\end{eqnarray}
The $K$ is the iteration number of this protocol. The total yield can be described as
\begin{eqnarray}
Y''=Y''_{1}+Y''_{1}+\cdots+Y''_{K}=\sum^{\infty}_{K=1}
Y''_{K}.\label{yield}
\end{eqnarray}
If the original state is the maximally entangled state with $a=b=\frac{1}{\sqrt{2}}$, we can get
$Y_{1}=0.25$, $Y'_{1}=0.5$, and $Y''=\frac{1}{2}+\frac{1}{4}+\cdots+\frac{1}{2^{K}}+\cdots=1$.

\begin{figure}[!h]
\begin{center}
\includegraphics[width=8cm,angle=0]{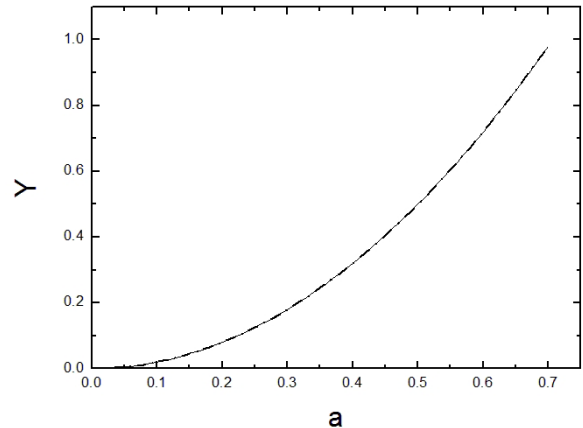}
\caption{The total entanglement yield $Y$ of getting a maximally HES after the concentration protocol being operated for K times in QND protocol. For numerical simulation, we choose $K=10$. It can be seen that the value of $Y$ largely depends on the initial coefficient $a$. When $a=\frac{1}{\sqrt{2}}$, $Y$ can reaches the maximum as 1.}
\end{center}
\end{figure}
We both calculated the entanglement yield in the QND protocol and VBS protocol in Fig. 8. It is shown that the yield  largely depends on the initial coefficient $a$.
The yield $Y'_{1}$ in single-photon protocol is essentially the case of $K=1$ in QND protocol. Interestingly, from Fig. 8, we only have
one curve.  Because by repeating the QND protocol for ten
times, the yield is consist with the VBS protocol. Therefore, it is shown that the VBS protocol is the optimal one. Actually,  the initial idea in this ECP is first proposed by the group of Deng \cite{hyperconcentration}.  The concentration step seems to be similar with Ref. \cite{hyperconcentration} for
we only need to operate the single polarized photon. Therefore, this ECP can also be extended to deal with the less-entangled state with
multi-photon and multi-coherent state. Only one of the parties who own the single particle needs to operate the protocol.

Finally, let us discuss some further realization in experiment. In
BS protocol, single-photon protocol and VBS protocol, we both resort to the linear
optics, which is feasible in current technology. In QND protocol, we
use the cross-Kerr nonlinearity to implement the QND. Though
cross-Kerr nonlinearity has been widely studied in quantum
information processes, we should acknowledge that it still has much
controversy. The main reason is that the largest natural cross-Kerr
nonlinearities are extremely
weak ($\chi^{3}\approx10^{-22}m^{2}V^{-2}$) \cite{kok1}. As mentioned
by Gea-Banacloche,   large shifts via the giant Kerr effect with
single-photon wave packet is impossible in current
technology \cite{Gea}.  Shapiro and Razavi also had the same results
with Gea-Banacloche\cite{Shapiro1,Shapiro2}. However, Hofmann
pointed out  that  a large phase-shift of $\pi$ can be achieved,
with a single two-level atom in a one-side cavity \cite{hofmann}. Fortunately, this protocol only
works for a small value of the cross-Kerr nonlinearity and it greatly decrease the experimental difficulty.
Actually, there are a great number of
works which focuses on   constructing the similar function of the QND for the photon-photon nonlinear interaction,
such as these based on quantum dot spins in microwave cavity \cite{cavity1,cavity2},  a cavity waveguide \cite{cavity waveguide},  hollow-core waveguides \cite{hollow-core waveguides}, a Rydberg atom ensemble \cite{rydberg}, and so on.

In summary, we have presented five different  ECPs for HES. All protocols can be used to achieve the tasks
of entanglement concentration. The second protocol can be seen as
the improvement of the first one and the third protocol can also be
regarded as the improvement of the first and the second protocol. These
protocols have several advantages: in the first protocol, it is
based on the simple optical elements, which can be realized in
current experiment. It does not need to know the exact
coefficients of the initial states.  The second protocol  is
also based on the linear optics and it is much simpler than the first
one. In the third protocol,   it only needs one pair of less-entangled
state, but can reach the same success probability as the first one.
Moreover, in the fourth protocol, the concentrated states can be
remained for further applications, by resorting to the QND. In
addition,  this ECP can be repeated to get a higher success
probability. The last ECP is the optimal one. By performing this ECP for one time, it can reach the success probability $2|a|^{2}$. Moreover,
 it based on the linear optics which makes it feasible in current experiment condition. All advantages will make these ECPs  useful in current
long-distance quantum communications.

\section*{ACKNOWLEDGEMENTS}
This work is supported by the National Natural Science Foundation of
China under Grant No. 11104159,  University Natural Science Research Project of Jiangsu Province
under Grant No. 13KJB140010,  Open Research
Fund Program of the State Key Laboratory of
Low-Dimensional Quantum Physics,  Tsinghua University, the open research fund of Key Lab of Broadband Wireless Communication and Sensor Network Technology, Nanjing University of Posts and Telecommunications, Ministry of Education (No. NYKL201303), and the Project
Funded by the Priority Academic Program Development of Jiangsu
Higher Education Institutions.

\end{document}